\documentclass[journal]{IEEEtran}
\usepackage{amsmath,amsfonts}
\usepackage{algorithmic}
\usepackage{algorithm}
\usepackage{array}
\usepackage[caption=false,font=normalsize,labelfont=sf,textfont=sf]{subfig}
\usepackage{textcomp}
\usepackage{stfloats}
\usepackage{url}
\usepackage{verbatim}
\usepackage{graphicx}
\usepackage{cite}
\usepackage{CJKutf8}
\usepackage{CJK}
\usepackage{multirow}
\usepackage{makecell}
\usepackage{tcolorbox}
\usepackage{soul}
\usepackage{threeparttable}

\usepackage{hyperref}
\begin{document}

\title{Physics-Informed Neural Optimal Control for Precision Immobilization Technique in Emergency Scenarios}

\author{
Yangye Jiang, Jiachen Wang, Daofei Li
\thanks{ \textcolor{black}{This work was supported by National Natural Science Foundation of China (Grant No.52372421) and the Zhejiang Key Laboratory of Intelligent Vehicle Comprehensive Safety. (Corresponding author: Daofei Li.)}
}
\thanks{The authors are with the Institute of Power Machinery and Vehicular Engineering, Zhejiang University and the Zhejiang Key Laboratory of Intelligent Vehicle
Comprehensive Safety, Hangzhou 310027, China. (e-mail: dfli@zju.edu.cn)}}


\markboth{IEEE TRANSACTIONS ON INTELLIGENT TRANSPORTATION SYSTEMS}{}


\maketitle

\begin{abstract}
Precision Immobilization Technique (PIT) is a potentially effective intervention maneuver for emergency out-of-control vehicle, but its automation is challenged by highly nonlinear collision dynamics, strict safety constraints, and real-time computation requirements. This work presents a PIT-oriented neural optimal-control framework built around PicoPINN (Planning-Informed Compact Physics-Informed Neural Network), a compact physics-informed surrogate obtained through knowledge distillation, hierarchical parameter clustering, and relation-matrix-based parameter reconstruction. A hierarchical neural-OCP (Optimal Control Problem) architecture is then developed, in which an upper virtual decision layer generates PIT decision packages under scenario constraints and a lower coupled-MPC (Model Predictive Control) layer executes interaction-aware control. To evaluate the framework, we construct a PIT Scenario Dataset and conduct surrogate-model comparison, planning-structure ablation, and multi-fidelity assessment from simulation to scaled by-wire vehicle tests. In simulation, adding the upper planning layer improves PIT success rate from 63.8\% to 76.7\%, and PicoPINN reduces the original PINN parameter count from 8965 to 812 and achieves the smallest average heading error among the learned surrogates (0.112 rad). Scaled vehicle experiments are further used as evidence of control feasibility, with 3 of 4 low-speed controllable-contact PIT trials achieving successful yaw reversal.
\end{abstract}


\section{Introduction}

\begin{figure*}[!t]
  \centering
  \includegraphics[width=0.82\textwidth]{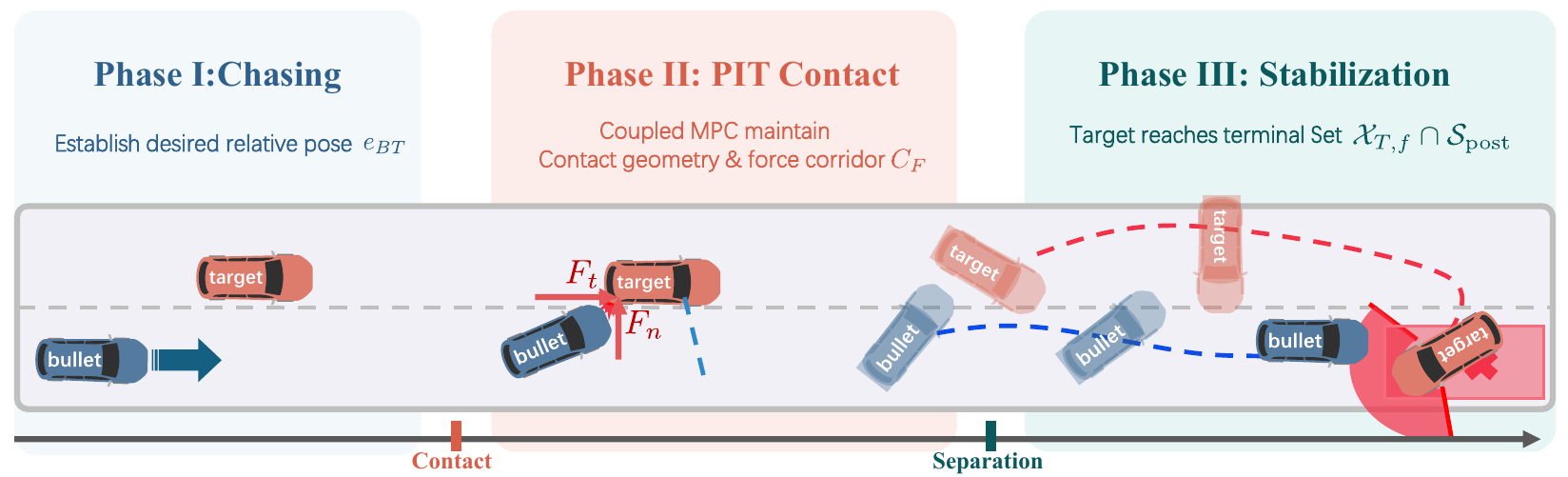}
  \caption{Schematic diagram of the PIT scenario.}
  \label{fig:PIT_scene}
\end{figure*}

\IEEEPARstart{W}{ith} the rapid development of intelligent transportation systems and autonomous driving, active intervention for safety-critical events has become an important research direction. In emergency intervention (stopping/rescue) scenarios, e.g. in Figure \ref{fig:PIT_scene}, Precision Immobilization Technique (PIT) with controlled rear-quarter contact is potentially to be used to guide the target vehicle toward a safe terminal state \cite{Zhou2008vehicle,Zhou2008Collision,Mascarenas2017Autonomous}. 

From a vehicle dynamics perspective, collision events involve tightly-coupled pre-impact, during-impact, and post-impact phases, where the short but highly nonlinear contact interval dominates downstream trajectory evolution \cite{Parseh2023Motion,Elkady2017Collision,Yang2025Post-impact,Yang2013Vehicle}. Such coupling makes PIT planning fundamentally different from standard lane-level autonomous driving tasks. Practical deployment requires terminal-state precision under strict constraints on road-boundary compliance, secondary-collision avoidance, and real-time computation.

On the other hand, the latest mid-to-high-end vehicles with driving automation functionalities are now commonly equipped with the sensing, onboard computing, and actuation foundations that are also required for automated PIT execution. However, a key bottleneck still remains in how to succeed PIT in constrained situations, especially while balancing physical fidelity and computational efficiency in control-oriented collision modeling. This motivates a PIT-specific framework that connects differentiable surrogate dynamics, neural optimal control, and safety-oriented evaluation under boundary operating conditions.


\subsection{Related work}
Collision dynamics modeling forms the theoretical foundation of PIT research. Early studies based on impulse--momentum principles provide interpretable analytical structure for light-impact prediction \cite{Brach1977Impact,Zhou2008Collision}. For detailed mechanism analysis, multibody and finite-element approaches improve fidelity and can reproduce deformation and force histories with high resolution \cite{Sousa2008Development,Chen2021deepa,Yildiz2012Multi-objective}. However, these methods are usually too computationally expensive for online planning and control.

For control-oriented applications, simplified impact-force representations remain widely used. Lumped spring-damper models and pulse-based approximations offer efficient formulations, with triangular or half-sine force profiles often adopted in low-to-moderate severity crash analysis \cite{Jonsen2009Identification,Ofochebe2015Performance,Iraeus2015Pulse,Huang2002,Wei2017Data-based,Huibers2001current,Kim2014Optimal}. Yet simplified templates alone are often insufficient when contact conditions vary significantly across scenarios.

Physics-Informed Neural Networks (PINNs) provide a promising alternative by embedding physical constraints into data-driven learning, thereby improving extrapolation and sample efficiency under limited data \cite{Raissi2019Physics-informed,Karniadakis2021Physics-informed}. Recent studies further emphasize that beyond adding physics residuals in the loss, network structure itself can be made physics-consistent through staged distillation and structure discovery. In particular, a physics-informed distillation framework has shown that decoupling physical regularization and parameter regularization, followed by clustering-based parameter reconstruction, can improve both efficiency and transferability in PDE(partial differential equation)-oriented tasks \cite{liu_2025_phi}.

In vehicle-related tasks, PINN variants have shown progress in car-following, trajectory prediction, and hybrid dynamics modeling \cite{Mo2021physics-informed,Shi2023Physics-informed,Geng2023physics-informed,Cheng2025hybrid,Fang2024Fine,Long2024Physics-informed}.

From the neural-OCP (Optimal Control Problem) perspective, recent work increasingly treats learned models as differentiable components within trajectory optimization rather than purely offline predictors. Representative directions include differentiable neural potential fields for collision-cost construction and toolchains that couple PyTorch models with numerical optimal-control solvers to preserve gradient flow in large-scale constrained optimization \cite{Alhaddad2023Neural,Salzmann2023Learning}. Related integrated planning-and-tracking studies further show that learned vehicle models can be embedded into a unified optimization loop to improve decision--execution consistency, especially under adhesion variation and emergency maneuvers \cite{Gao2024Integrated}.

From the neural-MPC (Model Predictive Control) perspective, most studies replace or augment prediction models within receding-horizon control using sequence networks (e.g., LSTM) or high-capacity neural dynamics. Existing results cover interactive traffic prediction--MPC coupling, car-following control in mixed traffic, and lateral-motion tracking with data-driven models. Recent real-time implementations further demonstrate that substantially larger neural dynamics models can run inside embedded gradient-based MPC loops \cite{Jeong2020Surround,Zhou2024Deep,Kim2024Data-Driven,Huang2023LSTM-MPC,Salzmann2023Real-time}. In summary, neural-OCP and neural-MPC share the goal of leveraging differentiable learned dynamics for constrained control, but the former emphasizes trajectory-level optimization structure, whereas the latter emphasizes online receding-horizon execution and real-time feasibility.

Nevertheless, most existing studies focus on non-contact traffic scenarios. Direct integration of collision-aware surrogate modeling with differentiable optimal control for PIT remains limited.
Building on our previous PINN-based collision modeling study \cite{Jiang2025PINN}, this work targets the planning-and-control layer and proposes a compact PIT-oriented surrogate (PicoPINN) for hierarchical neural optimal control and scenario-oriented evaluation.

\subsection{Main Contributions}
Despite important advances in collision modeling, physics-informed learning, and vehicle control, existing literature still lacks a PIT-oriented framework that jointly considers model compactness, differentiable optimization, and scenario-oriented evaluation. This work addresses these gaps through the following contributions:
\begin{itemize}
  \item A compact physics-informed surrogate model, denoted as the Planning-Informed Compact Physics-Informed Neural Network (PicoPINN), is developed by extending the original PINN with knowledge distillation, hierarchical agglomerative clustering, and parameter reconstruction. The resulting model preserves differentiability and physical consistency while reducing the parameter scale by approximately one order of magnitude, which is beneficial for neural-OCP deployment.
  \item A hierarchical neural-OCP formulation is established by embedding the lightweight surrogate into the PIT planning problem. The upper layer performs virtual PIT decision optimization, while the lower layer executes coupled MPC control that jointly optimizes actuation and collision interaction variables of the bullet vehicle, i.e., the controlled vehicle that initiates the PIT impact against the target vehicle.
  \item A PIT Scenario Dataset is constructed and used to benchmark model baselines and planning/control variants under boundary operating conditions, and is further validated through scaled vehicle experiments. This establishes a more application-relevant evaluation framework for controllable PIT.
\end{itemize}

The rest of this work is organized as follows. Section II first introduces the PIT scenario and then presents the vehicle-model layer used by planning and control. Section III develops the hierarchical neural optimal-control formulation. Section IV presents PIT Scenario Dataset construction and the simulation setup used for benchmarking. Section V reports simulation results and the scaled vehicle feasibility study. Finally, Section VI concludes this work and discusses future work.

\section{PIT Scenario and Vehicle Model}\label{sectionII}

Before introducing detailed model equations, the PIT operating scenario is first defined to make the interaction geometry and stage logic explicit. As shown in Figure \ref{fig:PIT_scene}, the bullet vehicle approaches the target vehicle under road-boundary and obstacle constraints, performs controllable side-rear contact during the PIT interaction, and then guides the target vehicle toward a safe terminal region. This scenario-level description is used consistently by the subsequent modeling and control formulation.

\subsection{Target-Vehicle PicoPINN Model}
\begin{figure*}[!t]
  \centering
  \includegraphics[width=0.85\textwidth]{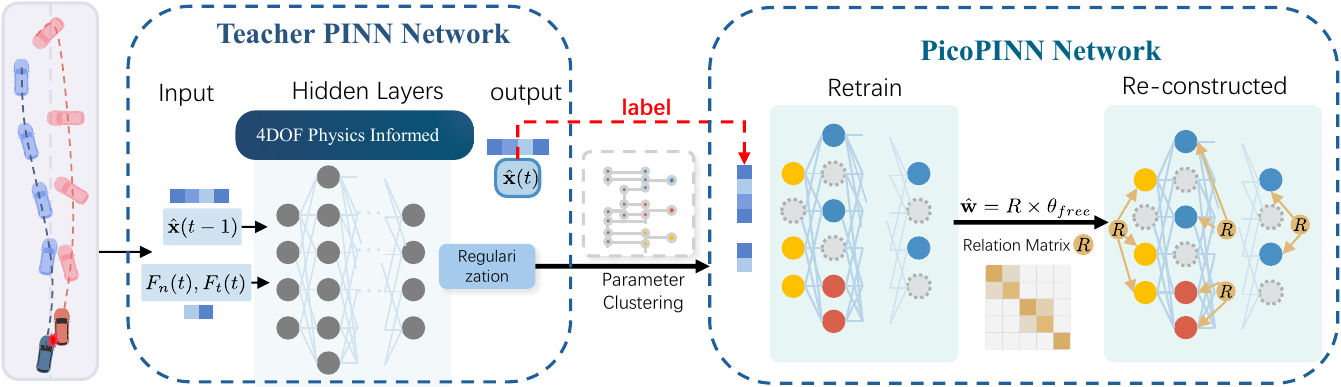}
  \caption{Schematic diagram of the PicoPINN architecture.}
  \label{fig:pico_pinn_architecture}
\end{figure*}

Traditional analytical models struggle to accurately capture highly nonlinear and discontinuous collision dynamics. To address this limitation while reducing repeated optimization cost, the original PINN surrogate \cite{Jiang2025PINN} is extended to a lightweight model termed PicoPINN (Planning-Informed Compact Physics-Informed Neural Network). As shown in Figure \ref{fig:pico_pinn_architecture}, PicoPINN inherits the physics-informed training paradigm and further introduces knowledge distillation, parameter clustering, and relation-matrix-based parameter reconstruction, making it more suitable as the predictive core of the subsequent neural-OCP formulation.

\begin{figure}[h]
  \centering
  \subfloat[Parameter clustering and resulting parameter reduction.\label{fig:param_clustering}]{
    \includegraphics[width=0.4\textwidth]{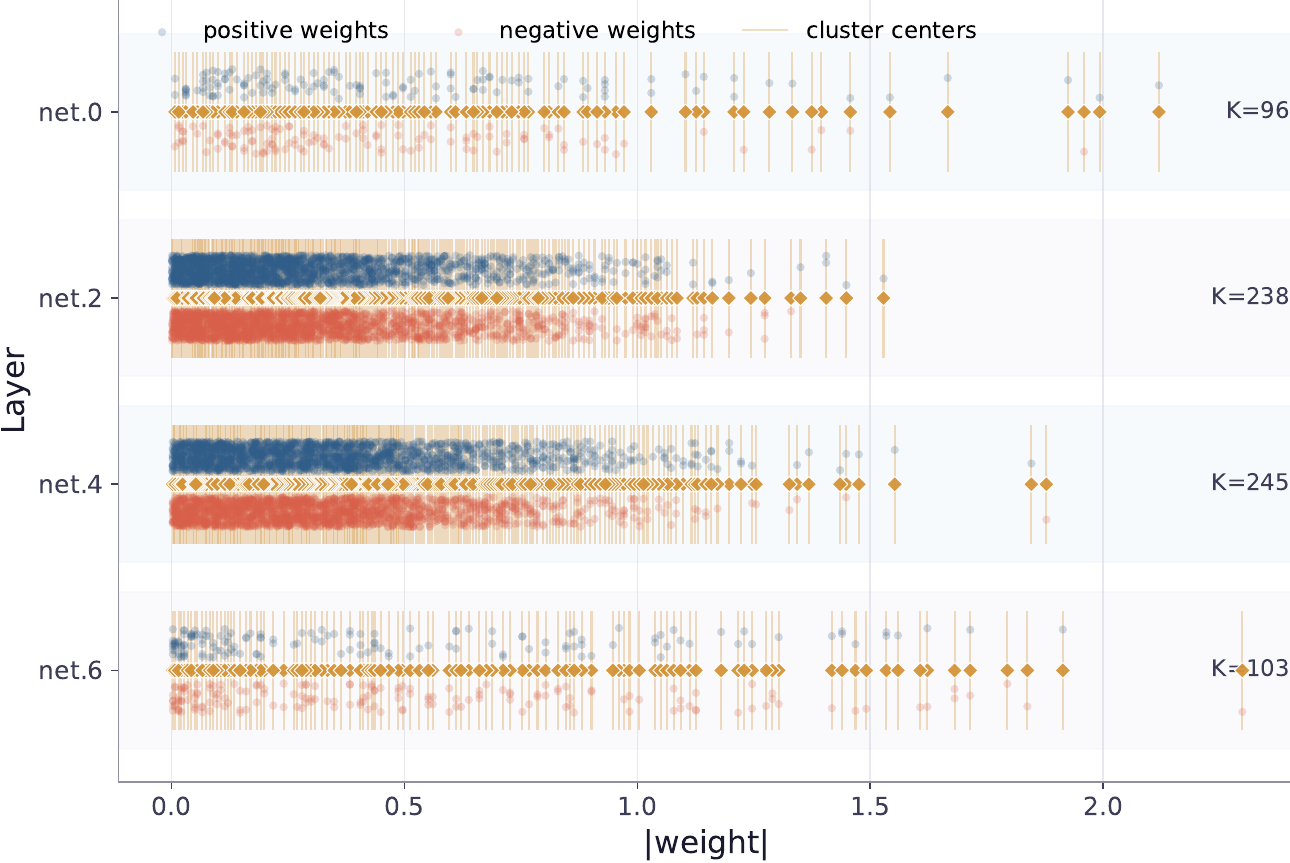}}
  \hfill
  \subfloat[Training curves of PicoPINN.\label{fig:pico_training_curves}]{
    \includegraphics[width=0.4\textwidth]{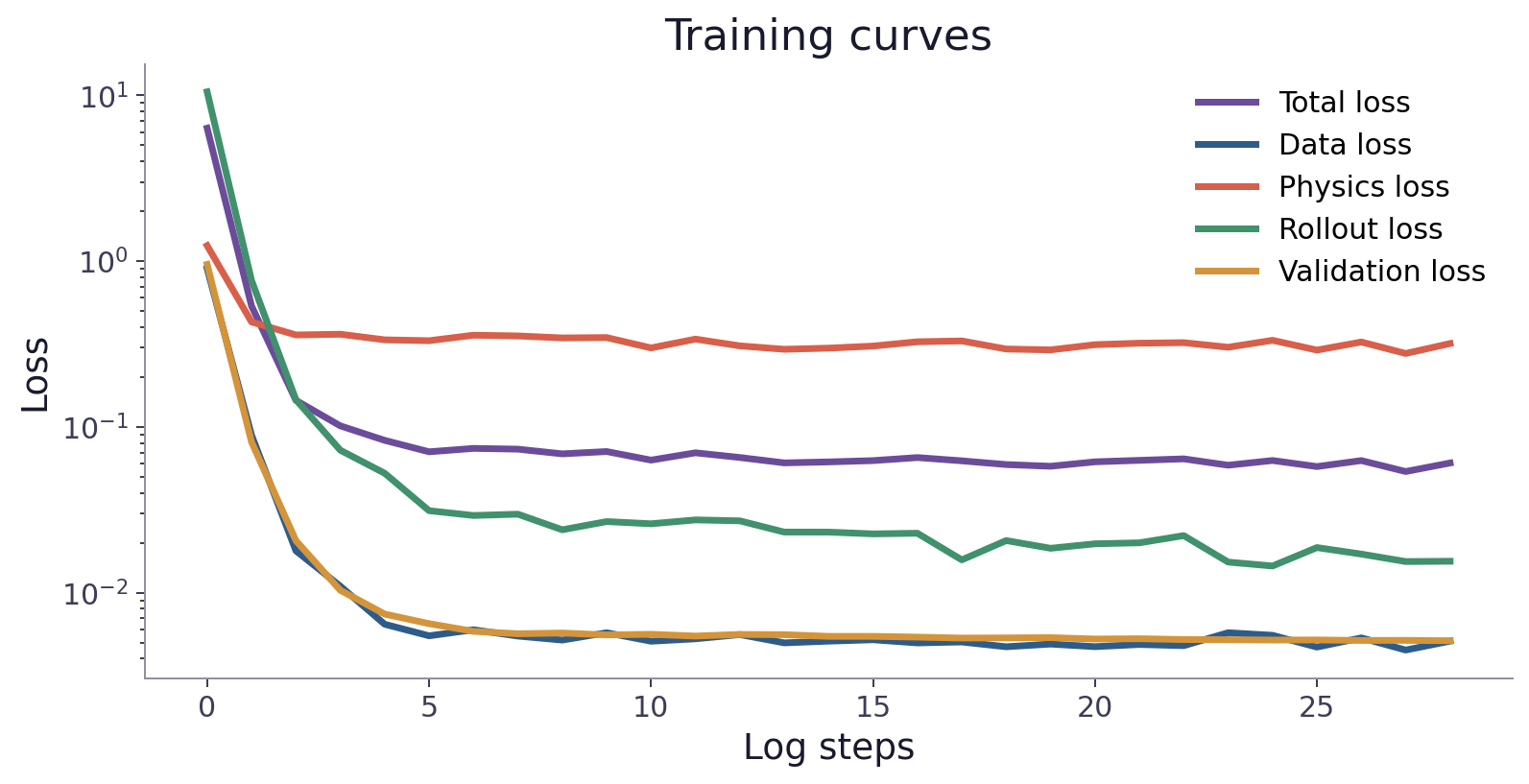}}
  \caption{Parameter clustering and training behavior of PicoPINN.}
  \label{fig:pico_clustering_training}
\end{figure}

Let the reconstructed parameter vector be denoted by $\boldsymbol{\theta}_{\mathrm{free}}$ and the corresponding relation matrix by $\mathbf{R}$. The compressed parameterization is written as
\begin{equation}
    \label{eq:phi_reconstruction}
    \hat{\mathbf{w}} = \mathbf{R}\boldsymbol{\theta}_{\mathrm{free}},
\end{equation}
where $\hat{\mathbf{w}}$ is the effective parameter vector used by the compact surrogate. In the present implementation, the teacher PINN network contains 8965 parameters, whereas PicoPINN reduces this number to approximately 800 trainable parameters after parameter clustering and reconstruction.
The clustering-induced parameter reduction and the training convergence/loss-evolution behavior are shown in Figure \ref{fig:pico_clustering_training}(a) and Figure \ref{fig:pico_clustering_training}(b), respectively.

The system dynamics are represented as
\begin{equation}
    \label{eq:vehicle_dynamics_pinn}
    \dot{\mathbf{x}}(t) = f_{\text{Pico}}(\mathbf{x}(t), \mathbf{u}(t), F_n(t), F_t(t); \boldsymbol{\omega}_{\mathrm{pico}})
\end{equation}
where $\mathbf{x} = [v_x, v_y, \psi, \dot{\psi}, \dot{\phi}, \phi, X, Y]^T \in \mathbb{R}^8$ is state vector, $ v_x $ and $ v_y $ are longitudinal and lateral velocities, $ \dot{\psi} $ is yaw rate, $ \dot{\phi} $ is roll rate, $ \phi $ is roll angle, $ (X, Y) $ is global position, and $ \psi $ is heading angle; $\mathbf{u} = [\delta, T_b]^T$ is the control input; $F_n$ and $F_t$ denote the contact normal and tangential components, respectively; and $\boldsymbol{\omega}_{pico}$ denotes the trained PicoPINN parameters.

The PicoPINN architecture maps the force-history-related input space to collision-sensitive state responses in the local contact frame:
\begin{equation}
    \label{eq:pinn_mapping}
    g_{\text{Pico}}(F_n(t), F_t(t), x(t-1)) \mapsto [x(t)]
\end{equation}

\noindent

The key advantage of PicoPINN lies in providing a continuously differentiable and computationally lighter dynamic surrogate throughout the collision maneuver.

\subsection{Bullet-Vehicle Control Model (Front Steering and Rear-Wheel Drive)}
For controller execution, the bullet vehicle is modeled by a control-oriented planar dynamics model with front-wheel steering and rear-wheel longitudinal drive. The bullet state and dynamic control input are defined as
\begin{equation}
    \label{eq:bullet_state_input}
    \begin{aligned}
    \mathbf{x}_{B}&=[X_B,Y_B,\psi_B,v_{x,B},v_{y,B},r_B]^T,\\
    \mathbf{u}_{B}^{\mathrm{dyn}}&=[\delta_{f,B},T_{r,B}]^T,
    \end{aligned}
\end{equation}
where $\delta_{f,B}$ is the front steering angle and $T_{r,B}$ is the rear-wheel drive/brake torque command.

The rear-wheel longitudinal force is generated from wheel torque as
\begin{equation}
    \label{eq:bullet_rear_force}
    F_{x,r,B}=\frac{\eta_d i_g}{R_w}T_{r,B},\qquad
    T_{r,B}\in[\underline{T}_{r},\overline{T}_{r}],\;\delta_{f,B}\in[\underline{\delta}_f,\overline{\delta}_f],
\end{equation}
with drivetrain efficiency $\eta_d$, gear ratio $i_g$, and effective wheel radius $R_w$.

Including contact-force terms in the bullet body frame, the continuous-time dynamics are
\begin{equation}
    \label{eq:bullet_ct_dyn}
    \begin{aligned}
    &\dot{X}_B = v_{x,B}\cos\psi_B-v_{y,B}\sin\psi_B,\\
    &\dot{Y}_B = v_{x,B}\sin\psi_B+v_{y,B}\cos\psi_B,\\
    &\dot{\psi}_B = r_B,\\
    &m_B(\dot{v}_{x,B}-r_Bv_{y,B}) = F_{x,r,B}-F_{y,f,B}\sin\delta_{f,B}\\
    &\hspace{1.5em} +F^{\mathrm{body}}_{x,c,B},\\
    &m_B(\dot{v}_{y,B}+r_Bv_{x,B}) = F_{y,f,B}\cos\delta_{f,B}+F_{y,r,B}\\
    &\hspace{1.5em} +F^{\mathrm{body}}_{y,c,B},\\
    &I_{z,B}\dot{r}_B = l_{f,B}F_{y,f,B}\cos\delta_{f,B}-l_{r,B}F_{y,r,B}\\
    &\hspace{1.5em} +M_{z,c,B},
    \end{aligned}
\end{equation}
where $F_{y,f,B}$ and $F_{y,r,B}$ are front/rear lateral tire forces and $F^{\mathrm{body}}_{x,c,B}$, $F^{\mathrm{body}}_{y,c,B}$, $M_{z,c,B}$ are interaction-induced terms from the contact pair.
The bullet vehicle model can be written in discrete form as
\begin{equation}
    \label{eq:bullet_dt_dyn}
    \mathbf{x}_{B,k+1}=f_B\big(\mathbf{x}_{B,k},\mathbf{u}_{B,k}^{\mathrm{dyn}},F^{\mathrm{body}}_{x,c,B}, F^{\mathrm{body}}_{y,c,B}\big).
\end{equation}

\section{Hierarchical Neural OCP Formulation}

\subsection{Hierarchical Neural OCP Statement}
\begin{figure*}[htbp]
    \centering
    \includegraphics[width=0.8\textwidth]{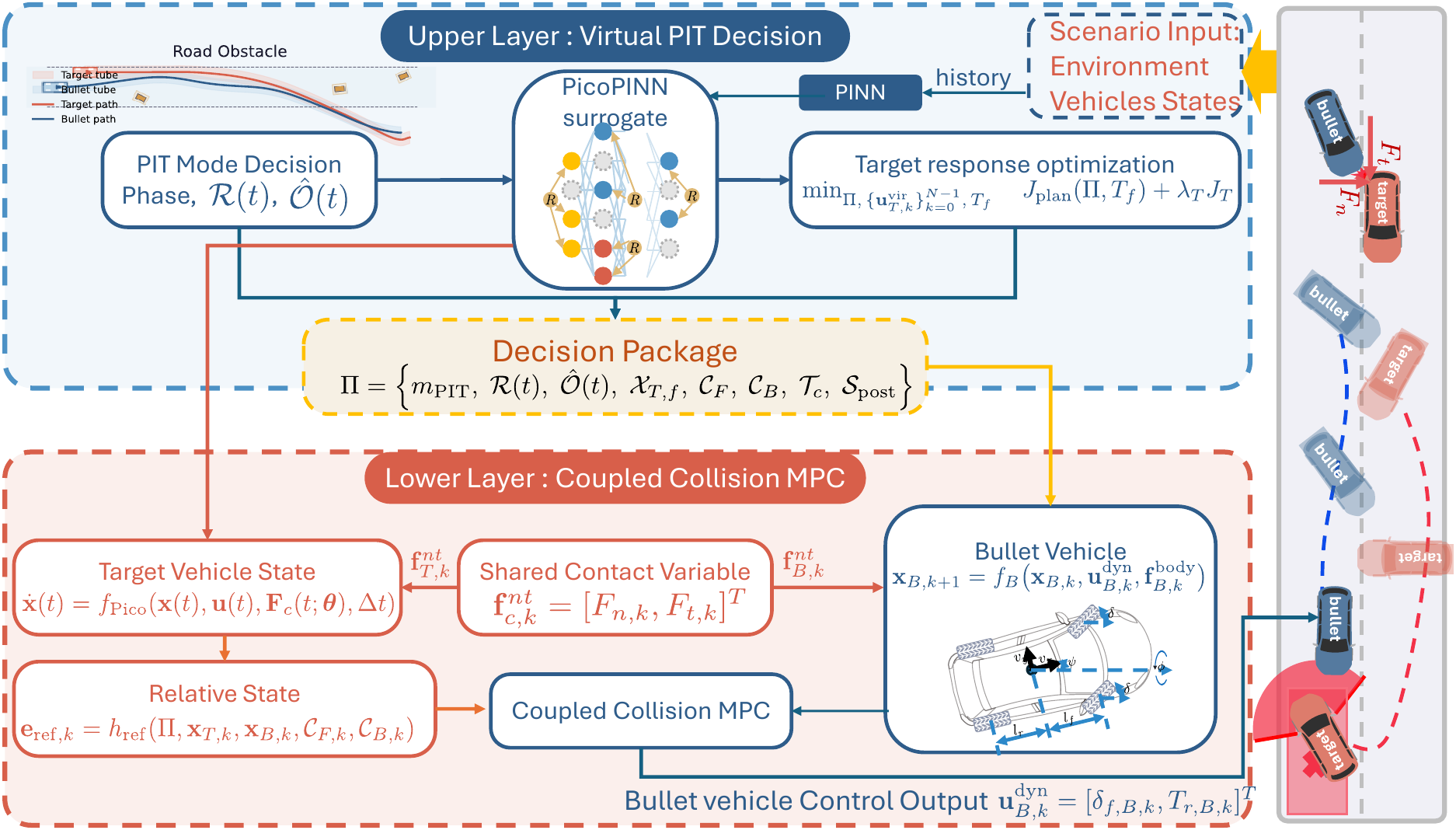}
    \caption{Decision-control framework of the proposed hierarchical neural OCP for PIT.}
    \label{fig:decision_control_framework}
\end{figure*}

The controllable PIT maneuver is formulated as a hierarchical problem composed of two layers. The upper layer performs virtual PIT decision optimization under road topology and obstacle constraints. Instead of directly actuating the target vehicle, it carries out a virtual PIT rollout over a 2~s prediction horizon and optimizes the terminal decision so that the target vehicle reaches a safe region with its yaw reversed relative to the lane direction. This upper-layer planner operates in a replanning manner every 1~s. The lower layer then executes the maneuver through a coupled collision MPC with a 1~s prediction horizon and a discretization step of 0.05~s. This distinction is important because the target vehicle is defined as the PIT object and is therefore not modeled as a directly actuated subsystem. The overall decision-control information flow is shown in Figure \ref{fig:decision_control_framework}.

\subsubsection{Virtual PIT Decision-Collision Layer}
Given the initial system state, the road geometry, and predicted surrounding obstacles, the upper layer jointly solves virtual PIT decision and virtual target-side collision inference. Its decision package is written as
\begin{equation}
    \label{eq:planning_output}
    \Pi = \left\{m_{\mathrm{PIT}},\;\mathcal{R}(t),\; \hat{\mathcal{O}}(t),\; \mathcal{X}_{T,f},\; \mathcal{C}_{F},\; \mathcal{C}_{B},\;\mathcal{T}_{c},\;\mathcal{S}_{\mathrm{post}}\right\},
\end{equation}
where $m_{\mathrm{PIT}}$ denotes the mode decision, $\mathcal{R}(t)$ denotes the road-boundary constraints, $\hat{\mathcal{O}}(t)$ denotes the predicted obstacle set, $\mathcal{X}_{T,f}$ denotes the desired terminal set of the target vehicle, $\mathcal{C}_{F}$ denotes the admissible $(F_n,F_t)$ force corridor, $\mathcal{C}_{B}$ denotes the bullet-vehicle tracking and contact envelope, $\mathcal{T}_{c}$ denotes the admissible contact-time window, and $\mathcal{S}_{\mathrm{post}}$ denotes the post-PIT safety set.

Within this upper layer, the target vehicle is represented by a virtual interaction descriptor
\begin{equation}
    \mathbf{u}_T^{\mathrm{vir}} = [F_n, F_t]^T,
\end{equation}
which denotes an interaction descriptor rather than a directly actuated command. A compact target-response objective is defined as
\begin{equation}
    \label{eq:target_mpc_cost}
    \begin{aligned}
    J_T =\;& w_{\psi}\left(\psi_{T,N}-\psi_{T,\mathrm{rev}}\right)^2 + w_v\|\mathbf{v}_{T,N}\|^2 \\
    &+ w_c\sum_{k=0}^{N-1}\mathrm{dist}\!\left(\mathbf{u}^{\mathrm{vir}}_{T,k},\mathcal{C}_{F,k}\right)^2,
    \end{aligned}
\end{equation}
where $\psi_{T,\mathrm{rev}}$ denotes the desired reversed heading and $\mathrm{dist}(\cdot,\mathcal{C}_{F,k})$ penalizes deviation from the feasible interaction corridor rather than enforcing an unrealistically exact force trajectory.

The upper-layer optimization can be summarized as
\begin{subequations}
\label{eq:planner}
\begin{align}
    \min_{\Pi,\,\{\mathbf{u}^{\mathrm{vir}}_{T,k}\}_{k=0}^{N-1},\,T_f} \quad & J_{\mathrm{plan}}(\Pi, T_f) + \lambda_T J_T \\
    \text{subject to} \quad & \dot{\mathbf{x}}(t)=f_{\text{Pico}}\big(\mathbf{x}(t), \mathbf{u}(t), \mathbf{u}^{\mathrm{vir}}_T(t); \boldsymbol{\omega}_{\mathrm{pico}}\big), \\
    & \mathbf{x}(0)=\mathbf{x}_0, \\
    & \mathbf{x}(t)\in \mathcal{R}(t)\setminus \hat{\mathcal{O}}(t), \\
    & \mathbf{u}^{\mathrm{vir}}_T(t)\in \mathcal{C}_{F}, \\
    & d\!\left(\mathbf{x}_T(t),\hat{\mathcal{O}}(t)\right)\ge d_{\mathrm{safe}}, \\
    & \mathbf{x}_T(T_f)\in \mathcal{X}_{T,f}\cap\mathcal{S}_{\mathrm{post}}.
\end{align}
\end{subequations}
where $J_{\mathrm{plan}}$ is the scenario-level decision objective and $\lambda_T$ is the coupling weight between virtual PIT decision and virtual target-response optimization.

\subsubsection{Coupled Collision MPC Control Layer}
The lower layer is formulated as a coupled MPC that jointly optimizes bullet actuation and collision interaction variables under the upper-layer decision package. The bullet dynamic actuation input is defined as
\begin{equation}
    \mathbf{u}_{B,k}^{\mathrm{dyn}}=[\delta_{f,B,k},T_{r,B,k}]^T,
\end{equation}
while the shared interaction variable in the contact frame is
\begin{equation}
    \mathbf{f}^{nt}_{c,k}=[F_{n,k},F_{t,k}]^T,
\end{equation}
with the interaction relationship
\begin{equation}
    \label{eq:force_pair}
    \mathbf{f}^{nt}_{T,k}=-\mathbf{f}^{nt}_{B,k}=\mathbf{f}^{nt}_{c,k}.
\end{equation}
Let $\gamma_k$ denote the contact-frame heading at step $k$. The contact force pair is mapped into each vehicle body frame as
\begin{equation}
    \label{eq:force_body_mapping}
    \begin{aligned}
    \mathbf{f}^{\mathrm{body}}_{T,k} &= \mathbf{R}(\psi_{T,k}-\gamma_k)\,\mathbf{f}^{nt}_{c,k},\\
    \mathbf{f}^{\mathrm{body}}_{B,k} &= -\mathbf{R}(\psi_{B,k}-\gamma_k)\,\mathbf{f}^{nt}_{c,k},
    \end{aligned}
\end{equation}
with $\mathbf{R}(\alpha)=\begin{bmatrix}\cos\alpha&-\sin\alpha\\\sin\alpha&\cos\alpha\end{bmatrix}$.

The lower-layer coupled MPC over horizon $N$ is summarized as
\begin{subequations}
\label{eq:coupled_mpc_problem}
\begin{align}
\min_{\{\mathbf{x}_{T,k},\mathbf{x}_{B,k},\mathbf{u}^{\mathrm{dyn}}_{B,k},\mathbf{f}^{nt}_{c,k}\}}\quad & J_{\mathrm{MPC}} \\
\text{subject to}\quad & \mathbf{x}_{T,k+1}=f_T(\mathbf{x}_{T,k},\mathbf{f}^{\mathrm{body}}_{T,k}), \\
& \mathbf{x}_{B,k+1}=f_B(\mathbf{x}_{B,k},\mathbf{u}^{\mathrm{dyn}}_{B,k},\mathbf{f}^{\mathrm{body}}_{B,k}), \\
& \mathbf{u}^{\mathrm{dyn}}_{B,k}\in\mathcal{U}_{B,k},\;\mathbf{f}^{nt}_{c,k}\in\mathcal{C}_{F,k}, \\
& \mathbf{e}_{BT,k}\in\mathcal{C}_{B,k}, \\
& \mathbf{x}_{T,k},\mathbf{x}_{B,k}\in \mathcal{R}_k\setminus\hat{\mathcal{O}}_k,
\end{align}
\end{subequations}
where $f_T(\cdot)$ is the PicoPINN-based target-vehicle transition model, $f_B(\cdot)$ is the front-steering/rear-drive bullet transition model in \eqref{eq:bullet_dt_dyn}, and $\mathbf{f}^{\mathrm{body}}_{T,k}$ and $\mathbf{f}^{\mathrm{body}}_{B,k}$ are transformed from the same optimized contact variable $\mathbf{f}^{nt}_{c,k}$.

Let the relative state between the two vehicles be denoted by
\begin{equation}
    \mathbf{e}_{BT} = [X_B-X_T,\; Y_B-Y_T,\; \psi_B-\psi_T,\; v_B-v_T]^T.
\end{equation}
The desired relative-contact state is generated online from the current interaction geometry and upper-layer corridors:
\begin{equation}
    \label{eq:eref_online}
    \mathbf{e}_{\mathrm{ref},k}=h_{\mathrm{ref}}\!\left(\Pi,\mathbf{x}_{T,k},\mathbf{x}_{B,k},\mathcal{C}_{F,k},\mathcal{C}_{B,k}\right),
\end{equation}
where $h_{\mathrm{ref}}(\cdot)$ returns a feasible relative-contact target satisfying the contact envelope $\mathcal{C}_{B,k}$ and consistency with the admissible force corridor $\mathcal{C}_{F,k}$.

The coupled-MPC objective can be written in a compact form as
\begin{equation}
    \label{eq:bullet_mpc_cost}
    \begin{aligned}
    J_{\mathrm{MPC}} =\;& w_TJ_T + w_e\sum_{k=0}^{N}\|\mathbf{e}_{BT,k}-\mathbf{e}_{\mathrm{ref},k}\|^2 \\
    &+ w_f\sum_{k=0}^{N-1}\mathrm{dist}\!\left(\mathbf{f}^{nt}_{c,k},\mathcal{C}_{F,k}\right)^2
    + w_u\sum_{k=0}^{N-1}\|\mathbf{u}^{\mathrm{dyn}}_{B,k}\|^2,
    \end{aligned}
\end{equation}
where $\mathbf{e}_{\mathrm{ref}}$ is the desired bullet-target relative-contact configuration. 


\subsubsection{Phase-Wise Interpretation}
The proposed formulation is intended for controllable-contact PIT. As shown in Figure \ref{fig:PIT_scene}, the maneuver is divided into three phases. \textbf{1) Chasing phase:} the bullet vehicle tracks the approach trajectory and establishes the desired relative pose. \textbf{2) PIT phase:} the coupled MPC maintains the contact geometry, timing window, and force or impulse corridor in a soft-tracking sense. \textbf{3) Stabilization phase:} after the PIT interaction, the predicted target motion is required to satisfy the terminal yaw-reversal and safety constraints while remaining free of secondary collision. The proposed lower layer is a practical controller for maintaining a controllable PIT interaction over a finite contact interval.

\subsubsection{Constraint Interface and Solver}
The upper layer summarizes mode, boundary, obstacle, terminal-set, contact-corridor, and safety constraints into a decision package and passes it to the lower-layer coupled MPC, while the l4casADi-based \cite{Salzmann2023Learning} implementation supports gradient-based optimization and receding-horizon execution.

\section{PIT Scenario Dataset Construction and Simulation Setup}
\subsection{PIT Scenario Dataset Construction}

The dataset is designed to characterize boundary operating conditions under which PIT is necessary, feasible, or unsafe while remaining consistent with real intervention cases. We first build a \emph{real-case prior pool} from publicly available PIT law-enforcement and crash videos, and then apply a CV-to-BEV \cite{yang2024depthv2} extraction pipeline to obtain key descriptors, including initial speed, relative pose, road topology, and obstacle context. One representative frame per scene is further processed for coarse semantic tagging via LLM (large language model) (e.g., road type, approximate adhesion level, lane count and width, and lead-vehicle class).

The scenario space is defined by the joint variation of video-inferred initial states, road geometry, and environment-related factors. Representative categories include:
\begin{itemize}
    \item straight-road, curved-road, and lane-boundary-constrained PIT geometries;
    \item low-, medium-, and high-adhesion conditions;
    \item sparse-to-dense obstacle-context configurations near roadway boundaries;
\end{itemize}

\subsubsection{Data Acquisition, Screening, and Annotation}
The dataset is constructed through CV-based real-case extraction, conditional simulation generation, and quality control. Video-derived priors and coarse semantic tags are used to generate distribution-consistent scene-schema parameters. Each case is executed with vehicle-type-specific CarSim \texttt{.simfile} models selected by scene features, with collision-force inputs from the matched FEA force library in our previous work \cite{Jiang2025PINN}; Carla is used for visualization and replay-based curation. Infeasible or trivial samples are removed by rule-based pre-screening and manual review.

\begin{figure*}[htbp]
    \centering
    \includegraphics[width=0.9\textwidth]{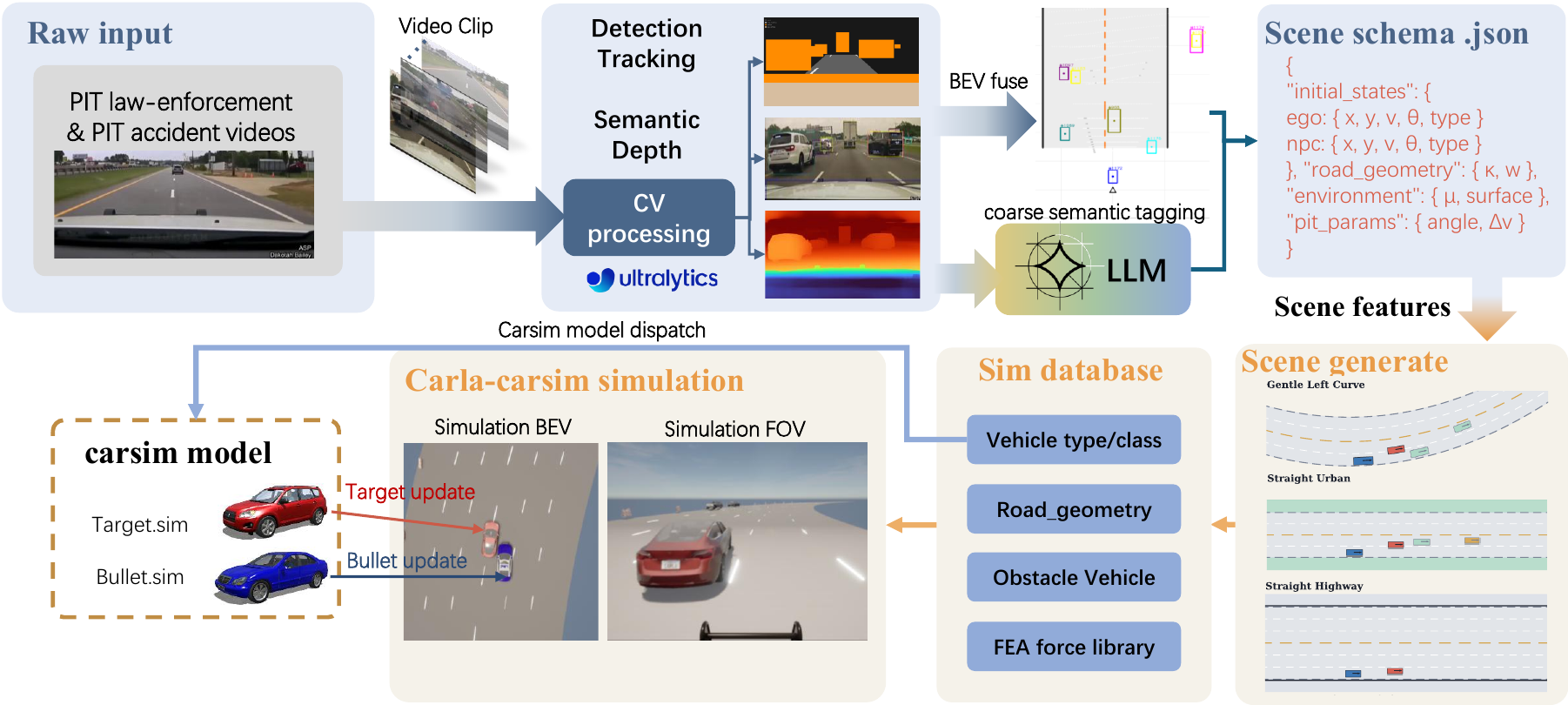}
    \caption{Dataset-construction and simulation-scene-generation pipeline from PIT videos to the CarSim--Carla PIT Scenario Dataset.}
    \label{fig:dataset_pipeline}
\end{figure*}

The overall dataset-construction and simulation-scene-generation workflow is shown in Fig. \ref{fig:dataset_pipeline}, and key dataset statistics are summarized in Table \ref{tab:database_statistics}.

\begin{table}[htbp]
\centering
\caption{Statistics of the PIT Scenario Dataset}
\label{tab:database_statistics}
\begin{tabular}{lc}
\hline
\textbf{Item} & \textbf{Value} \\
\hline
Source video events & 214 \\
CV-extracted valid clips & 156 \\
Valid parameterized priors & 117 \\
Generated scenarios & 3600 \\
Retained realistic PIT scenes & 2048 \\
\hline
\end{tabular}
\end{table}

\begin{figure*}[htbp]
    \centering
    \subfloat[Category breakdown of generated simulation dataset.\label{fig:category_breakdown}] {
        \includegraphics[width=0.45\textwidth]{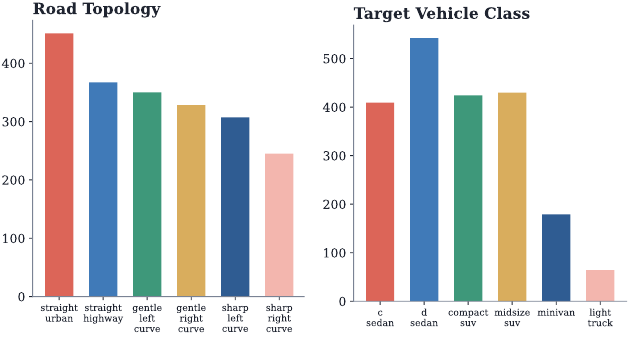}}
    \hfill
    \subfloat[KL-divergence comparison between generated and real datasets.\label{fig:kl_divergence_comparison}] {
        \includegraphics[width=0.45\textwidth]{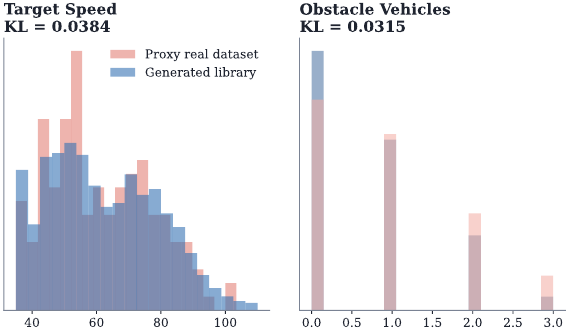}}
    \caption{Distribution-oriented evaluation of dataset generation quality.}
    \label{fig:dataset_generation_distribution}
\end{figure*}

The generated-scenario composition and its distributional consistency with the real-case prior are illustrated in Fig. \ref{fig:dataset_generation_distribution}(a) and Fig. \ref{fig:dataset_generation_distribution}(b), respectively. Fig. \ref{fig:dataset_generation_distribution}(a) reports category-level composition, and Fig. \ref{fig:dataset_generation_distribution}(b) compares KL divergence between generated and real datasets across representative attributes.

The retained scenes are distributed across straight-urban (22.0\%), straight-highway (17.9\%), and curved-road (60.1\%) layouts.

\subsection{Comparative Experiment Design for the Proposed PIT Strategy}

\subsubsection{Benchmark Baselines}
Because no widely accepted public benchmark is dedicated to PIT planning and control, the comparison is organized in two levels. The first is \emph{surrogate-model comparison}, used to assess whether PicoPINN is suitable for downstream neural OCP; representative baselines are the conventional 4DOF analytical model, the original PINN surrogate. The second is \emph{planning-structure comparison} under the same PicoPINN dynamics, namely full hierarchical neural OCP (with the upper planning layer) versus direct coupled MPC (without the upper planning layer). This two-level design avoids redundant combinations and isolates whether improvements come from surrogate compactness or upper-layer planning.

\subsubsection{Evaluation Metrics}
The evaluation metrics are grouped into two sets. \emph{Surrogate-model metrics} quantify prediction quality and deployment-oriented closed-loop behavior, including RMSE error for $\phi$; overall PIT success rate under each surrogate; success-rate breakdowns across road topologies and adhesion levels; planning and control runtimes. \emph{Planning-and-control metrics} quantify task completion quality and real-time feasibility, including PIT success rate (primary metric), terminal heading error, minimum safety margin, and planning/control runtimes. Unless otherwise stated, the non-percentage scalar metrics reported in the integrated benchmark are averages over the evaluated runs.
For reproducible evaluation, a PIT trial is counted as successful if and only if the target vehicle reaches the prescribed terminal condition within the allowed horizon $T_{\max}$ while satisfying safety constraints throughout the maneuver.  Specifically, the run must satisfy $|e_{\psi}(T)|\leq \varepsilon_{\psi}$ and remain free of boundary violation, unsafe rebound, and secondary collision. This definition is used consistently for simulation and scaled vehicle tests.

\section{Simulation Results and Scaled Vehicle Experiments}

\subsection{Integrated Benchmark Results}

\begin{table*}[htbp]
\centering
\caption{Integrated Benchmark of Planning Structure, Surrogate Models, and Success/Failure Composition}
\label{tab:integrated_benchmark}
\setlength{\tabcolsep}{3pt}
\renewcommand{\arraystretch}{1.12}
\begin{tabular}{p{3.15cm}cccccc}
\hline
\makecell[l]{\textbf{Planning}\\\textbf{Structure}} & \makecell{\textbf{Succ.}\\\textbf{(\%)}} & \makecell{\textbf{Head.Err.}\\\textbf{(rad)}} & \makecell{\textbf{Min Margin}\\\textbf{(m)}} & \makecell{\textbf{Bound. Violation}\\\textbf{(\%)}} & \makecell{\textbf{2nd Coll.}\\\textbf{(\%)}} & \makecell{\textbf{Plan/Ctrl}\\\textbf{(ms)}} \\
\hline
\makecell[l]{PicoPINN + upper-plan\\+ cMPC} & 76.7 & 0.106 & 0.84 & 1.9 & 0.8 & 842 / 24.6 \\
\makecell[l]{PicoPINN + cMPC\\(without upper layer)} & 63.8 & 0.114 & 0.62 & 6.8 & 4.5 & 0 / 35.9 \\
\hline
\end{tabular}

\vspace{0.6em}

{\scriptsize
\setlength{\tabcolsep}{1.8pt}
\begin{tabular}{lccccccccccccccc}
\hline
\multirow{2}{*}{\textbf{Surrogate Model}} & \multirow{2}{*}{\makecell{\textbf{Succ.}\\\textbf{(\%)}}} & \multicolumn{2}{c}{\textbf{Road (\%)}} & \multicolumn{3}{c}{\textbf{Adhesion (\%)}} & \multirow{2}{*}{\makecell{\textbf{Head.Err.}\\\textbf{(rad)}}} & \multicolumn{3}{c}{\textbf{Runtime / Size}} & \multicolumn{5}{c}{\textbf{Failure Cause (\%)}} \\
\cline{3-4}\cline{5-7}\cline{9-11}\cline{12-16}
 &  & \textbf{Straight} & \textbf{Curved} & \textbf{Low-$\mu$} & \textbf{Mid-$\mu$} & \textbf{High-$\mu$} &  & \textbf{Plan} & \textbf{Ctrl} & \textbf{Param.} & \makecell{\textbf{Decision}\\\textbf{Infeas. in Chasing}} & \makecell{\textbf{No-Term.}\\\textbf{by $T_{\max}$}} & \makecell{\textbf{Bound.}\\\textbf{Violation}} & \textbf{2nd Coll.} & \makecell{\textbf{Unsafe}\\\textbf{Rebound}} \\
\hline
4DOF analytical model & 58.9 & 65.4 & 54.7 & 42.3 & 59.8 & 71.5 & 0.161 & 2310 & 107.4 & -- & 13.2 & 13.8 & 7.7 & 3.7 & 2.7 \\
Original PINN & 74.9 & 79.1 & 72.0 & 64.2 & 75.6 & 81.3 & 0.118 & 1346 & 47.8 & 8965 & 12.9 & 5.8 & 3.5 & 1.7 & 1.2 \\
PicoPINN & 76.7 & 80.3 & 74.2 & 66.1 & 77.8 & 82.6 & 0.112 & 842 & 24.6 & 812 & 12.7 & 5.1 & 3.0 & 1.4 & 1.1 \\
\hline
\end{tabular}
}
\end{table*}

Table \ref{tab:integrated_benchmark} summarizes planning-structure and surrogate-model performance, including success rates, heading error, safety metrics, runtime, model size, and failure-cause composition. Except for model parameter count, the remaining scalar metrics are reported as averages over the evaluated runs.

Introducing the upper planning layer increases PIT success from 63.8\% to 76.7\% (+12.9 percentage points), increases the minimum safety margin from 0.62 m to 0.84 m, and reduces the average heading error from 0.114 rad to 0.106 rad, while reducing boundary-violation and secondary-collision rates from 6.8\% to 1.9\% and from 4.5\% to 0.8\%, respectively. Among surrogate models, PicoPINN provides the best overall trade-off: highest overall success rate (76.7\%), best low-adhesion success (66.1\%), and smallest average heading error (0.112 rad), while reducing parameter count from 8965 (Original PINN) to 812 (90.9\% reduction). The additional failure type, early exit due to infeasible PIT decision in the chasing phase, remains near 13\% for all models (12.7\%--13.2\%), indicating that most gains are reflected in reduced boundary violation, secondary collision, and unsafe rebound.

\begin{figure*}[htbp]
    \centering
    \includegraphics[width=0.7\textwidth]{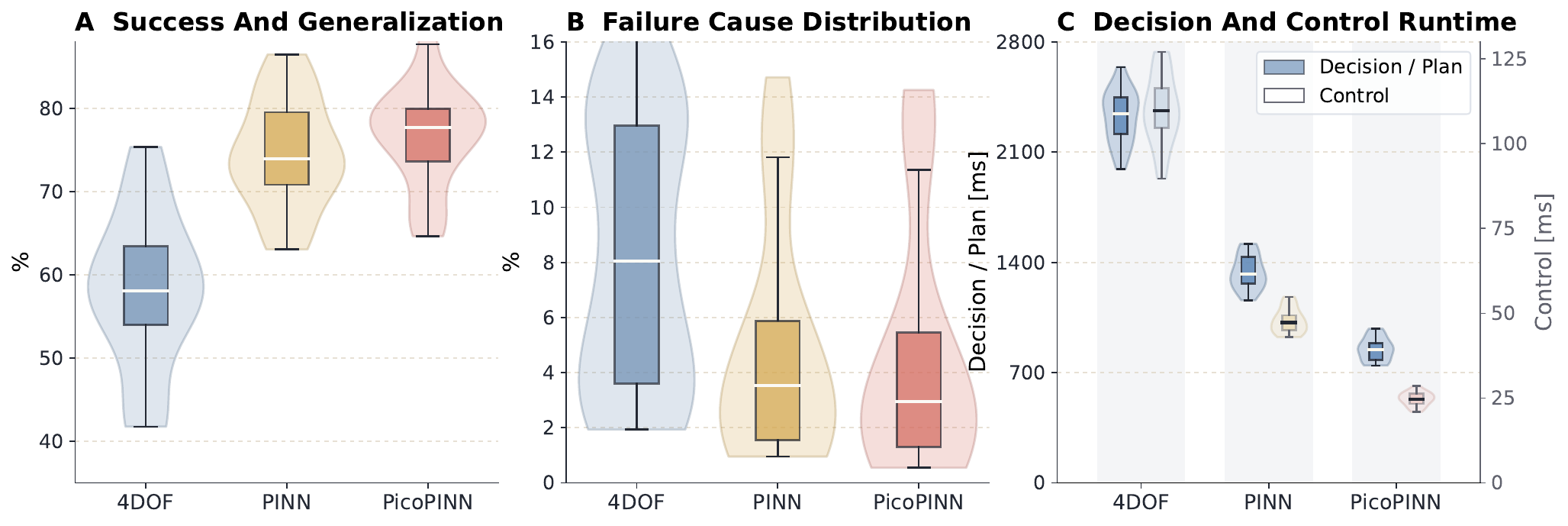}
    \caption{Visualization-oriented benchmark summary consistent with Table \ref{tab:integrated_benchmark}.}
    \label{fig:benchmark_visualization}
\end{figure*}

Figure \ref{fig:benchmark_visualization} supplements Table \ref{tab:integrated_benchmark} for qualitative interpretation. Panel A visualizes mission/road/adhesion success indicators; Panel B visualizes failure-cause composition; and Panel C jointly visualizes planning and control runtimes, showing that planning remains substantially more expensive than control for all models.

\subsection{Representative Simulation Case Analysis}

\begin{figure*}[htbp]
    \centering
    \includegraphics[width=0.98\textwidth]{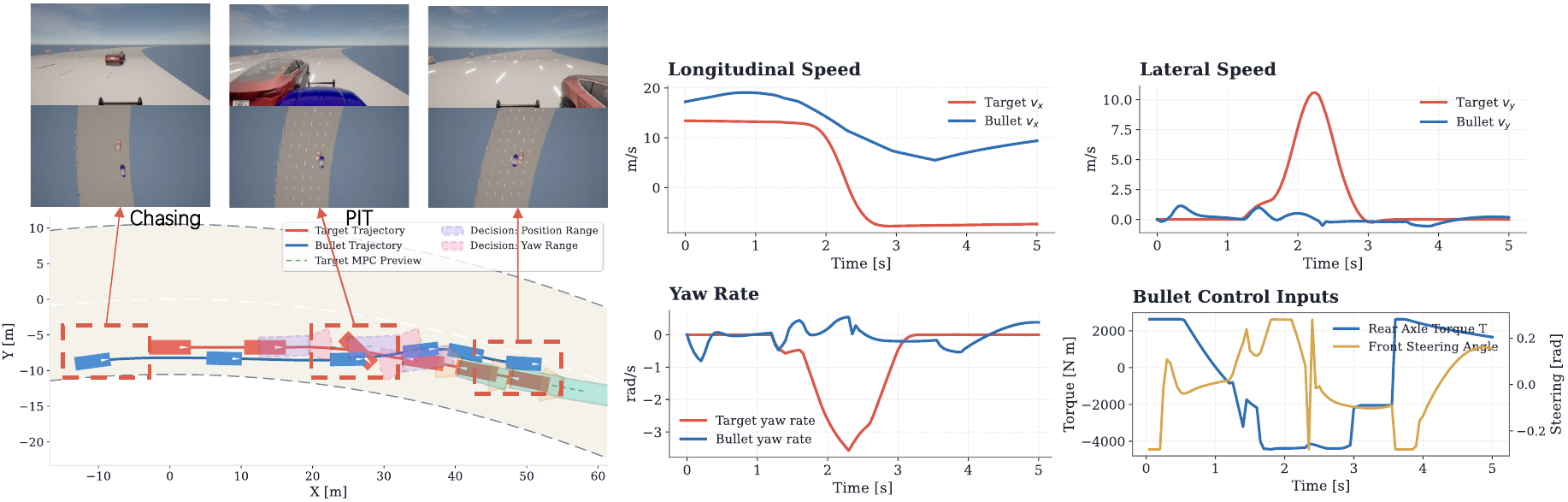}
    \caption{Representative successful case: trajectory evolution, state responses, and bullet control inputs.}
    \label{fig:case_success}
\end{figure*}

\begin{figure*}[htbp]
    \centering
    \includegraphics[width=0.98\textwidth]{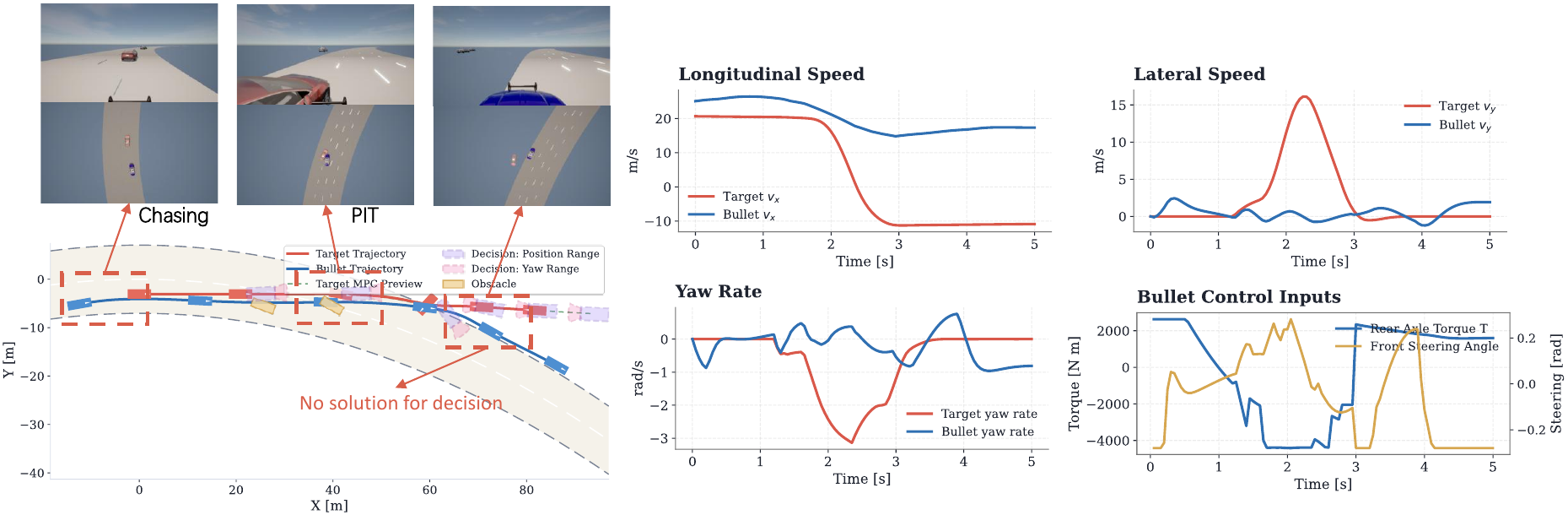}
    \caption{Representative failed case: trajectory evolution, state responses, and bullet control inputs.}
    \label{fig:case_fail}
\end{figure*}

To complement aggregate metrics, two representative cases are compared through scenario settings and trajectory-level responses: a successful case and a failed case, shown in Figures \ref{fig:case_success} and \ref{fig:case_fail}, respectively.

In the successful case, the road is dry asphalt with high adhesion ($\mu=0.9214$), no surrounding dynamic obstacles, and relatively strong PIT excitation (contact angle $21.196^\circ$, speed difference $3.804$ m/s). The target yaw-rate peak generated during contact is followed by stable recovery, lateral-speed response remains bounded, and bullet control inputs recover from saturation in time; therefore, the terminal trajectory stays inside the desired corridor.

In the failed case, the road is wet asphalt ($\mu=0.5812$), two forward obstacles are present, and PIT excitation is weaker (contact angle $15.546^\circ$, speed difference $2.936$ m/s). Absolute speeds are higher (ego/target: $23.954/21.018$ m/s versus $17.232/13.429$ m/s), and the initial longitudinal gap is larger (7.462 m versus 5.424 m). These factors jointly reduce effective lateral impulse and leave less post-contact correction margin. Correspondingly, the failed case shows a larger lateral-speed peak, longer post-contact drift, and longer near-saturated rear-axle torque and steering periods, causing the terminal state to leave the desired set within the finite horizon (boundary violation).

\subsection{Scaled Vehicle Feasibility Study}

\subsubsection{Experimental Platform}
\begin{figure}[htbp]
    \centering
    \includegraphics[width=0.5\textwidth]{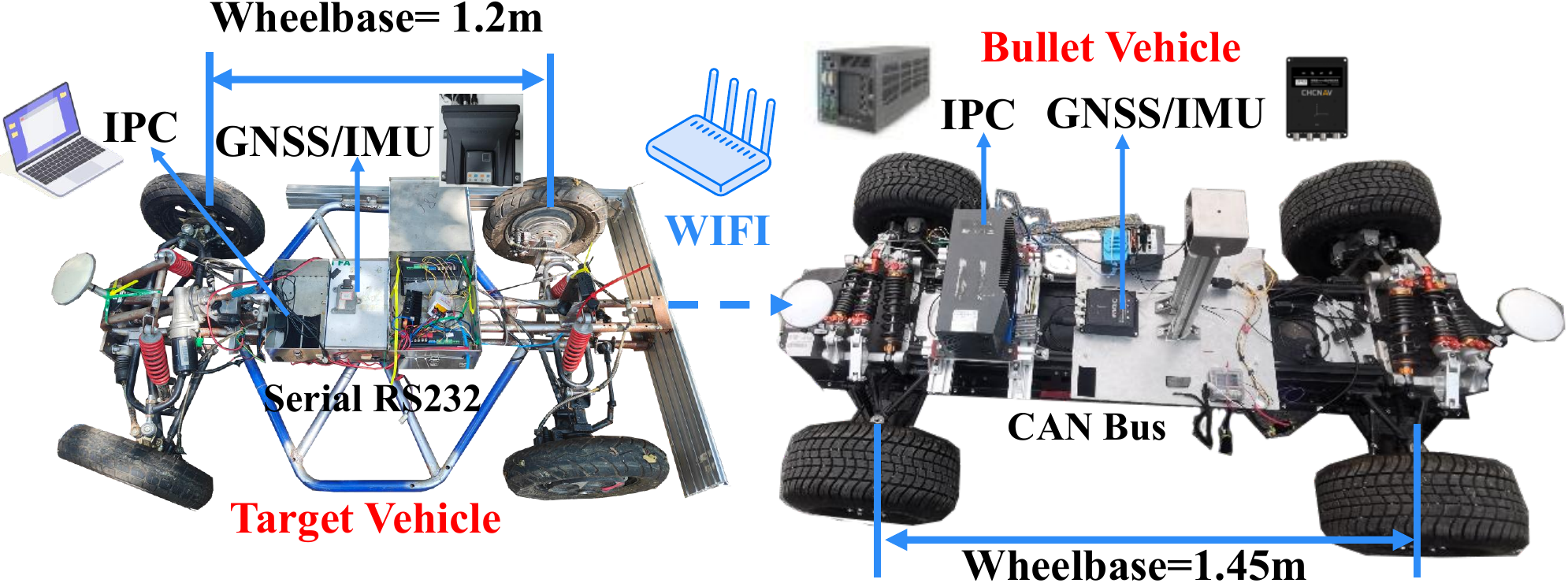}
    \caption{Half-scaled experiment platforms: bullet and target vehicles.}
    \label{fig:real_vehicle}
\end{figure}

To further evaluate practical feasibility, tests on half-scaled by-wire vehicles are conducted. As shown in Figure \ref{fig:real_vehicle}, the experiment platform consists of a bullet vehicle and a target vehicle, with an approximate wheelbase-based scaling ratio of 1:2 relative to full-size passenger cars. The bullet vehicle has dimensions of approximately $2100\times1120\times455$ mm, a mass of approximately $290$ kg, front-wheel steering, and independent four-wheel drive, whereas the target vehicle has dimensions of approximately $1900\times900\times350$ mm, a mass of approximately $100$ kg, front-wheel steering, and rear-wheel drive. Both vehicles are equipped with pneumatic tires so that the nonlinear vehicle responses relevant to PIT can still be reproduced at the scaled level. Here, the platform is used to examine low-speed controllable-contact control feasibility rather than to reproduce a full-scale crash event.

Each vehicle is equipped with an industrial personal computer for onboard planning and control, and the state-estimation system is built on GNSS-IMU sensing. According to the previously verified platform specification, the position accuracy is approximately $0.01$ m and the angular-rate accuracy is approximately $0.002$ rad/s. Inter-vehicle communication is realized through a local wireless network, whereas low-level commands are transmitted through the CAN bus.

\subsubsection{Test Procedure}
The scaled vehicle experiment follows the same three-phase logic as the proposed control framework. In the chasing phase, the bullet vehicle establishes the desired relative pose with respect to the target vehicle. In the PIT phase, the planned contact corridor is maintained by the coupled MPC so that the target vehicle is driven toward yaw reversal while remaining inside the road boundary and avoiding secondary collision. In the stabilization phase, the resulting target motion is observed to determine whether the required terminal yaw condition is reached without unsafe rebound or secondary contact. The results of scaled vehicle experiments are shown in Table \ref{tab:real_vehicle}. Except for the binary success label, here the remaining scalar metrics are reported as averages in the `Overall` row.

\begin{table*}[htbp]
\centering
\caption{Summary of Real-Vehicle PIT Validation Results}
\label{tab:real_vehicle}
\begin{tabular}{lccccc}
\hline
\textbf{Case} & \textbf{Target Speed (km/h)} & \textbf{Relative Impact Speed (km/h)} & \textbf{Impact Angle (deg)} & \textbf{Heading Error (rad)} & \textbf{Remark} \\
\hline
R1 & 7.8 & 3.1 & 21.4 & 0.182 & Successful \\
R2 & 8.2 & 4.2 & 24.8 & 0.209 & Successful \\
R3 & 7.6 & 2.6 & 18.9 & 0.074 & Successful \\
R4 & 8.4 & 0.8 & 10.1 & 1.127 & Failed \\
Overall & 8.0 & 2.7 & 18.8 & 0.398 & 3/4 successful \\
\hline
\end{tabular}
\end{table*}

\begin{figure*}[htbp]
    \centering
    \includegraphics[width=0.9\textwidth]{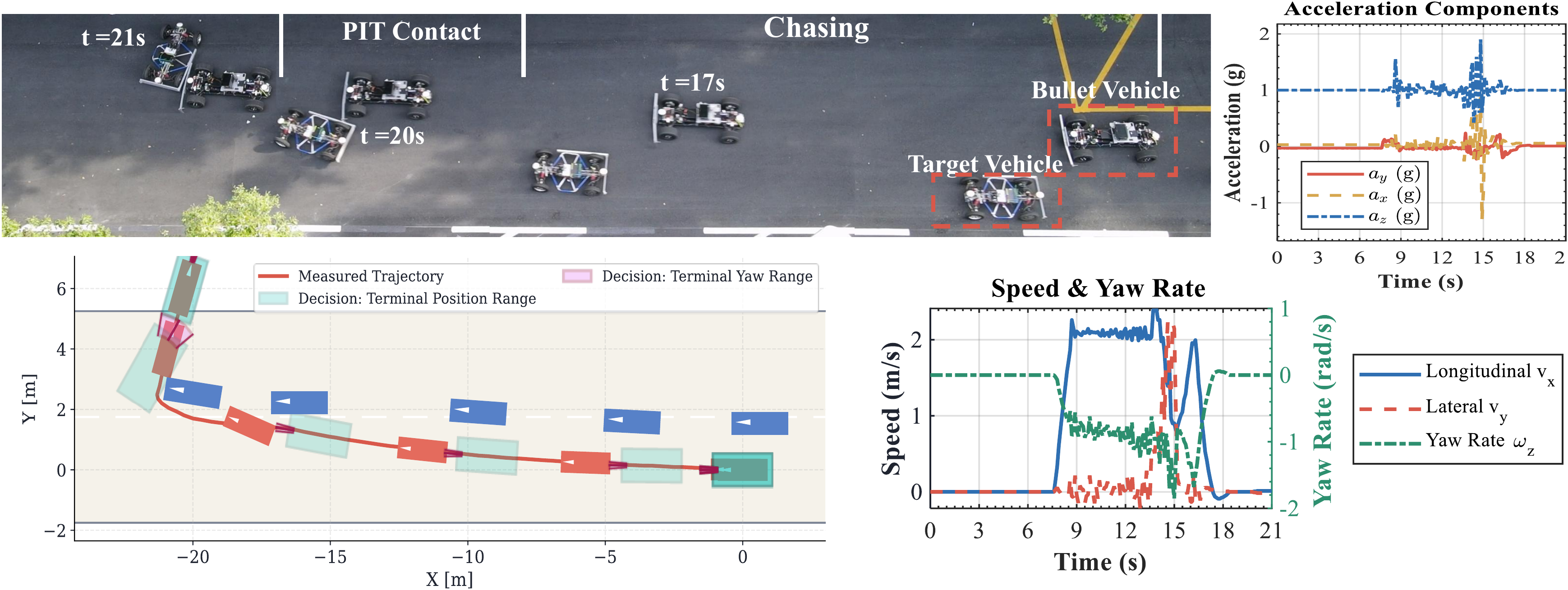}
    \caption{Representative scaled vehicle test Case 1 (R1).}
    \label{fig:exp_set1}
\end{figure*}

As a representative vehicle test run, Case 1 (R1) is shown in Figure \ref{fig:exp_set1}. Under a target speed of 7.8 km/h and a relative impact speed of 3.1 km/h, the maneuver achieves the required terminal yaw condition with a heading error of 0.182 rad, consistent with the successful label in Table \ref{tab:real_vehicle}. This result provides preliminary physical evidence that the proposed controller can maintain controllable contact and guide the target toward the intended post-impact evolution in the scaled vehicle setting.

\subsubsection{Discussion}
First, the half-scaled vehicle tests should be interpreted as evidence of control feasibility rather than full-size vehicle deployment validation, since the latter involves much higher costs of resources. The main purpose here is to examine whether the proposed upper-planning and coupled-MPC framework can maintain the intended contact evolution, achieve yaw reversal, and remain operable under sensing noise, communication latency, and actuation uncertainty. With this, the 3/4 success rate provides preliminary support that the controller can coordinate contact timing and post-impact stabilization in a physical closed loop. However, further full-scale deployment readiness is to be validated in future work.

Second, the failed scaled case is consistent with the failure trends observed in simulation. In Table \ref{tab:real_vehicle}, R4 corresponds to the smallest relative impact speed (0.8 km/h) and the smallest impact angle (10.1$^\circ$), which together imply insufficient lateral excitation for reliable yaw reversal. This observation is aligned with the simulation-side failure analysis, where weak PIT excitation and limited post-contact correction margin lead to boundary-related failure. The consistency between simulated failure mechanisms and the scaled physical result is encouraging, because it suggests that the controller is capturing the right qualitative sensitivity to contact severity even though exact post-impact trajectories still differ across fidelities.

Third, a non-negligible sim-to-real gap remains. The controller relies on a compact differentiable surrogate and simplified corridor-based interaction description, whereas the physical platform still contains unmodeled effects such as sensing bias, communication delay, tire nonlinearities, compliance, and scale-dependent contact behavior. Accordingly, the present study should be understood as a staged validation pipeline: high-fidelity simulation provides the main comparative evidence, and the scaled tests provide a practical check that the control architecture remains executable on hardware. Extending the physical evidence to a broader set of scaled conditions, incorporating explicit delay or uncertainty adaptation, and pursuing larger-scale experimental validation are therefore important next steps before any real-world deployment can be considered.

\section{Conclusion}
This work develops a PIT-oriented neural optimal-control framework based on a lightweight physics-informed surrogate model. By extending the previous PINN with knowledge distillation, hierarchical clustering, and parameter reconstruction, PicoPINN provides a compact differentiable dynamics surrogate for neural-OCP-based PIT planning. The framework formulates PIT as a hierarchical problem: an upper virtual PIT decision layer plans target-side response and constraints, while a lower coupled-MPC layer executes the maneuver under road and obstacle constraints across chasing, PIT, and stabilization phases. The study also shifts evaluation from isolated case validation to scenario-oriented assessment by introducing a PIT Scenario Dataset, a benchmark framework for surrogate and control comparisons, and a multi-fidelity validation pipeline spanning high-fidelity simulation and preliminary scaled vehicle control-feasibility experiments.

Future work will focus on expanding benchmark scale and quantitatively identifying which classes of PIT boundary scenarios benefit most from the proposed method in terms of safety and consistency.


\bibliographystyle{IEEEtran}
\bibliography{IEEEabrv,ref}

\vspace{10pt}
\bf{Author info}\vspace{-33pt}
\begin{IEEEbiography}[{\includegraphics[width=1in,height=1.25in,clip,keepaspectratio]{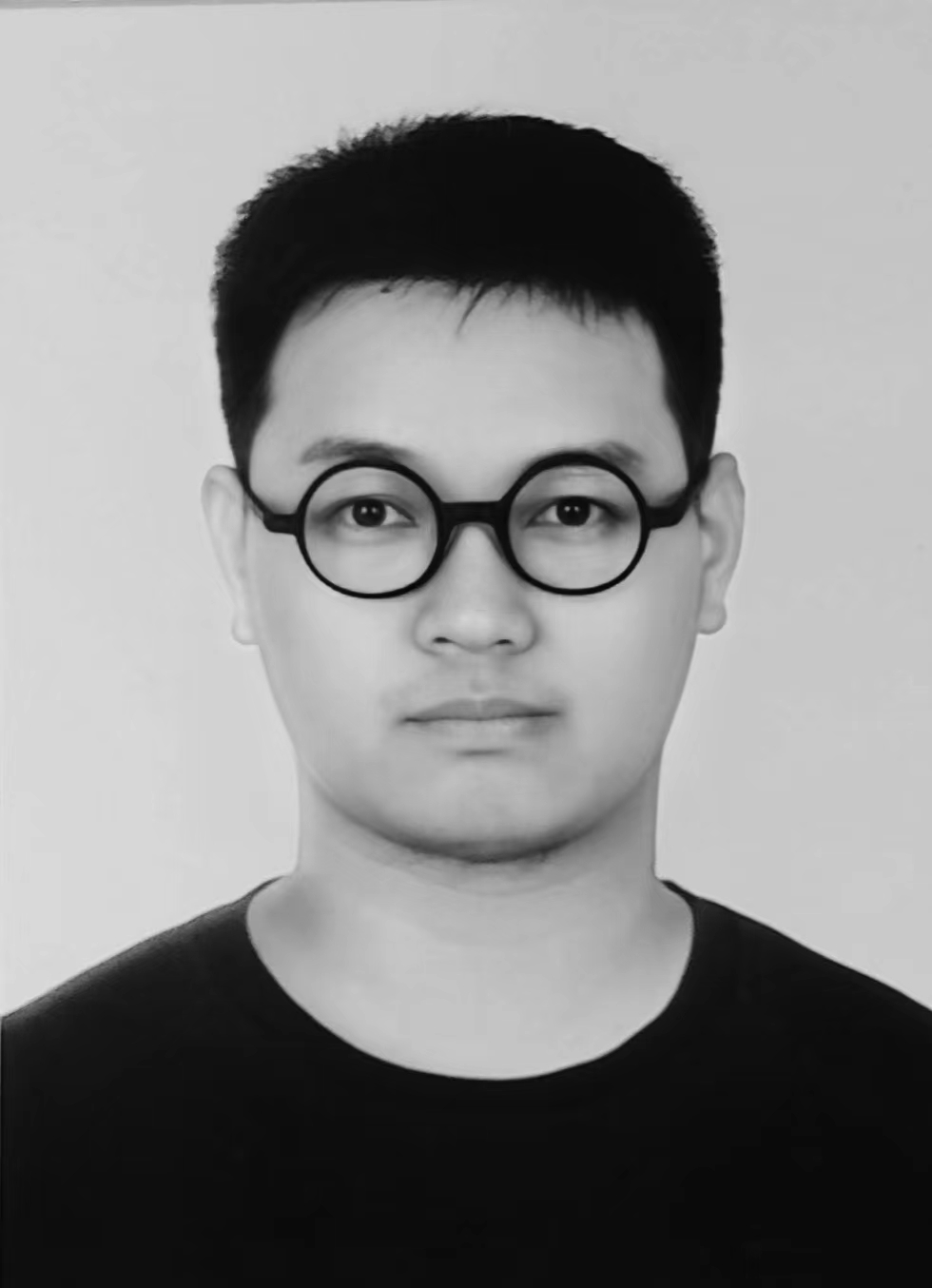}}]{Yangye Jiang}
obtained his B.S. degree of Vehicle Engineering from Zhejiang University, Hangzhou, China, in 2023. 
He joined the Research Group of Human-Mobility-Automation, ZJU in August 2022 and is currently pursuing for his Ph.D. of Power Engineering and Engineering Thermophysics. His research interests include automated driving decision and eco-driving.
\end{IEEEbiography}

\begin{IEEEbiography}[{\includegraphics[width=1in,height=1.25in,clip,keepaspectratio]{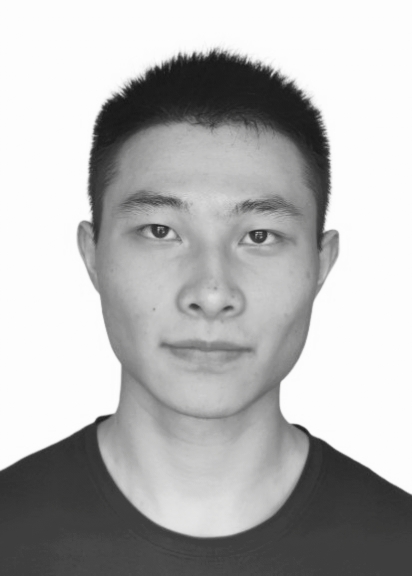}}]{Jiachen Wang}
obtained his B.S. degree of Vehicle Engineering from Zhejiang University, Hangzhou, China, in 2024. 
He joined the Research Group of Human-Mobility-Automation, ZJU in March 2023 and is currently pursuing a Master's degree in Power Engineering and Engineering Thermophysics. His research interests include automated driving decision and motion sickness.
\end{IEEEbiography}

\begin{IEEEbiography}[{\includegraphics[width=1in,height=1.25in,clip,keepaspectratio]{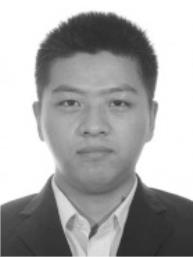}}]{Daofei Li}
  received the B.S. degree in Vehicle Engineering from the Jilin University, Changchun, China, in 2003, 
  and the Ph.D. degree in Vehicle Engineering from the Shanghai Jiao Tong University, Shanghai, China, in 2008.
  Since June 2008, he joined the Institute for Power Machinery and Vehicular Engineering, Faculty of Engineering, Zhejiang University (ZJU), Hangzhou, China. He was a Visiting Scholar with the University of Missouri-Columbia in 2011, and later with the University of Michigan, Ann Arbor, Michigan from 2014 to 2016. He is currently Associate Professor with ZJU and directs the Research Group of Human-Mobility-Automation. His research interests include vehicle dynamics and control, driver model, motion sickness and autonomous driving. He is a member of IEEE and the IEEE Intelligent Transportation Systems Society.
\end{IEEEbiography}

\vfill

\end{document}